\documentclass{conm-p-l}

\usepackage{graphicx}

\usepackage{}

\newtheorem{theorem}{Theorem}[section]

\theoremstyle{definition}

\theoremstyle{remark}

\numberwithin{equation}{section}

\begin{document}

\title[Predicting critical transitions with cubical homology]{Towards the prediction of critical transitions in spatially extended populations with cubical homology.}


\author[L.S. Storch]{Laura S. Storch}
\address{Department of Mathematics \\ William \& Mary \\ Williamsburg, VA}
\email{laura.s.storch@gmail.com}

\author[S.L. Day]{Sarah L. Day}
\address{Department of Mathematics \\ William \& Mary \\ Williamsburg, VA}
\email{sldayx@wm.edu}

\subjclass[2010]{92(Primary) and 54(Secondary)}

\date{}

\begin{abstract}
The prediction of critical transitions, such as extinction events, is vitally important to preserving vulnerable populations in the face of a rapidly changing climate and continuously increasing human resource usage.  Predicting such events in spatially distributed populations is challenging because of the high dimensionality of the system and the complexity of the system dynamics.  Here, we reduce the dimensionality of the problem by quantifying spatial patterns via Betti numbers ($\beta_0$ and $\beta_1$), which count particular topological features in a topological space.  Spatial patterns representing regions occupied by the population are analyzed in a coupled patch population model with Ricker map growth and nearest-neighbors dispersal on a two-dimensional lattice.  
We illustrate how Betti numbers can be used to characterize spatial patterns by type, which in turn may be used to track spatiotemporal changes via Betti number time series and characterize asymptotic dynamics of the model parameter space.  En route to a global extinction event, we find that the Betti number time series of a population exhibits characteristic changes.  We hope these preliminary results will be used to aide in the prediction of critical transitions in spatially extended systems.  Additional applications of this technique include analysis of spatial data (e.g., GIS) and model validation.  
\end{abstract}

\maketitle

\section{Introduction}
The dynamics of spatially distributed populations are difficult to understand and predict, both due to their complex behavior and high dimensionality.  Improving our understanding of these systems, and attempting to predict critical changes in their behavior, become vitally important in the context of a globally changing climate, habitat destruction, and exploitation pressures, all of which contribute to increased dynamical volatility and extinction risk (e.g., \cite{McLaughlin02}, \cite{Hsieh06}, \cite{Anderson08}).  Here, we aim to understand and predict the dynamics of a spatially distributed population by examining spatial patterns and how they change over time.  We accomplish this by analyzing the topological features of spatial patterns representing the regions of space that are occupied by members of the population.  
More specifically, using cubical homology to calculate the first and second Betti numbers ($\beta_0$ and $\beta_1$) of the union of occupied patches in a two-dimensional lattice model allows us to quantify properties of these population patterns.  These numbers, $\beta_0$ and $\beta_1$, count the number of connected components and one-dimensional holes, respectively, in the population pattern.  Population time series are generated via a coupled patch model with Ricker map growth (\cite{Ricker54}), symmetric nearest-neighbors dispersal, and absorbing boundaries.  We find that $\beta_0$ and $\beta_1$ can be used to characterize spatial patterns by type, and to track spatial pattern changes over time.  We track population global extinction events via Betti number time series and find characteristic changes in $\beta_0$ and $\beta_1$ en route to global extinction, suggesting that Betti numbers may be useful in the prediction of critical transitions.  

Coupled patch lattice models and tools from computational topology have a separate, established history in the literature.  Coupled patch models are relatively simple, yet can exhibit complicated dynamics in both space and time.  Studies have experimented with the coupling of two patches (e.g., \cite{Sole92}, \cite{Hastings93}, \cite{Wysham08}), the coupling of multiple patches along a line (e.g., \cite{Chate88}, \cite{Kaneko86}, \cite{Kaneko90}, \cite{Kaneko92}, \cite{Labra03}, \cite{Willeboordse03}, \cite{White05}) and the coupling of patches in a two-dimensional spatial lattice (e.g., \cite{Kaneko89}).  Here, we focus on the two-dimensional lattice models.  Likewise, computational topology has been used to measure structure in many systems, including time-varying models and data.  These studies range from using cubical homology to study patterns in models (\cite{Krishan07}, \cite{Cochran13}, \cite{Day09}) to using simplicial persistent homology to study time evolving patterns of population point cloud data (\cite{Corcoran17}, \cite{Topaz15}, \cite{Botnan18}, \cite{Buchin13}, \cite{Sumpter10}).   Our focus here is to use cubical homology to quantify and track population patterns representing the occupied regions of space, forming the foundation for later studies using the more sophisticated tool of cubical persistent homology to study the additional information offered by patch-wise abundance data. 


Our goal is to use topology to study the complex patterns that arise in coupled patch lattice models.  
In a similar model to our own, Kaneko (\cite{Kaneko89}) finds that a population distributed across two-dimensional space can fall into several typical patterns, depending on the parameter combination.  Such patterns include checkerboard, fully-developed spatiotemporal chaos, and the frozen random pattern, which consists of a population pattern that appears largely periodic in space and time, with small sections of sustained non-periodic behavior.  The Kaneko model consists of a coupled patch model with logistic map growth, symmetric nearest-neighbors dispersal, and periodic boundary conditions.  Our model employs a different growth map and boundary conditions, but we observe qualitatively similar dynamics.  While previous work has also focused on characterizing system dynamics over the model parameter space, here, we focus specifically on the topological features of the spatial patterns produced by the model.  The topological features allow for the characterization of spatial patterns by type, allow for observation of changes in spatial patterns over time, and provide an additional tool for labeling of the parameter space.  

One goal of our spatial pattern analysis is to determine if there is a characteristic change in spatial patterns that occurs en route to a critical dynamical transition, and if we can use this information to predict critical transitions.  More specifically, we focus on the changes in spatial pattern en route to a global extinction event.  To our knowledge, this is a novel approach to the prediction of critical transitions, which currently focuses on dynamical or statistical markers of impending change.   

Scheffer et al. (\cite{Scheffer09}) outline several potential symptoms of critical dynamical transitions, which manifest as critical slowing down events.  The symptoms of an impending transition can include slowing of recovery from perturbation, increased variance, and increased autocorrelation.  Such symptoms have been observed in real-world systems (e.g., \cite{Hsieh06}) but direct applications remain limited.  In spatially extended systems, the distribution of the population may also provide early warning signals, e.g., increase in spatial coherence preceding an event (\cite{Scheffer09}), or self-organized patchiness (\cite{Rietkerk04}).   

For spatially extended systems specifically, increased attention has been given to the prediction of critical transitions in semiarid grassland environments.  These tend to self-organize into characteristic patterns, dependent on rainfall (e.g., \cite{Scheffer09}, \cite{Rietkerk04}, \cite{Gowda14}, \cite{Gowda16}).  The characteristic pattern of the system corresponds with the dynamical state, and so tracking changes in pattern may provide early warning signals.  Grassland models transition through the standard “gaps--labyrinth--spots” sequence (Figure 1) with decreasing rainfall.  When they reach the patchy “spots” phase, small regions of grass are isolated in the environment and dynamically decoupled, and a catastrophic shift may be imminent (\cite{Rietkerk04}).    

\begin{figure}
\centering
\includegraphics[scale = 0.35]{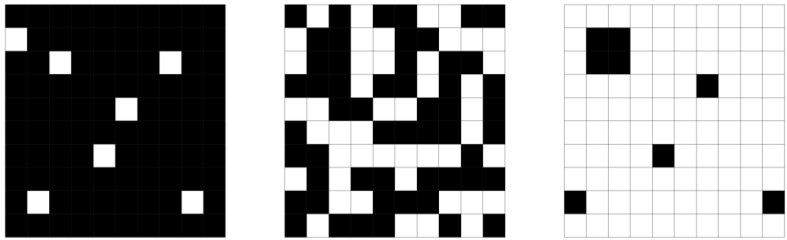}
\caption{Example population patterns on a $10\times 10$ lattice, displaying gaps, labyrinth, and spots, respectively, moving from left to right.  Black color indicates that the patch is occupied, white color indicates that the patch is unoccupied.  We compute Betti numbers for the pattern given by the union of the occupied (black) patches.  Please note that each patch is a closed square so that two edge or corner adjacent patches form a connected pattern.  Patch boundaries are displayed for visualization purposes.  The leftmost image has $\beta_0 =1$ and $\beta_1 =6$, the middle image has Betti numbers $\beta_0 =3$ and $\beta_1 =3$, the rightmost image has $\beta_0 =5$ and $\beta_1 =0$.}
\label{fig:GLS}
\end{figure}

Here, we employ a topological approach to quantifying population spatial patterns and tracking their changes over time.  We focus on two types of spatial features of the populations: connected components and holes, counted by the first and second Betti numbers.  Figure \ref{fig:GLS} illustrates the Betti numbers of example population patterns, showing the gaps, labyrinth, and spots phases, respectively.  The images are binary and each two-dimensional patch is categorized as either occupied (black) or unoccupied (white).  We compute Betti numbers for the patterns given by the union of the black patches.  Since each patch is considered to be a closed square, two edge or corner adjacent patches overlap and their union forms a connected set.   

Analyzing a high dimensional system via Betti numbers has several advantages.  First, the usage of Betti numbers reduces the dimensionality of the system to two coarse-grain measurements, $\beta_0$ and $\beta_1$.  Although we reduce the dimensionality of the system, we are retaining information about the spatial patterns, which provides different insights than tracking, e.g., total population.  Lastly, we can also use this technique for a wide variety of additional applications, such as the analysis of datasets and/or model validation.  
For example, Chung \& Day (\cite{Chung18}) obtain topological measurements of three-dimensional firn data (the stage between snow and ice), which are used to accurately depict the “pores” in the firn (one- and two-dimensional holes) to inform gas transfer models.  We also envision using this technique on GIS satellite images to track changes in, e.g., vegetation cover (additionally, see \cite{Krishan07, Gameiro05} for applications of cubical homology and Betti numbers to the study of pattern formation and evolution in convection and phase separation models).  

\section{Background}

\subsection{Population model}
We use a density-dependent coupled patch population model, where growth and dispersal occur on a 2-dimensional $N\times N$ lattice.  Patch-wise population abundances are recorded in an $N\times N$ matrix $X$.  That is, for $1\leq i,j \leq N$, $X(i,j)$ is the population abundance in patch $(i,j)$.  The growth phase occurs first, where the population in each individual patch reproduces, independently of population values in other patches.  A set portion of the population in each patch is then dispersed uniformly amongst the four nearest (co-dimension one) neighboring patches.  

\begin{figure}
    \centering
    \includegraphics[scale = 0.15]{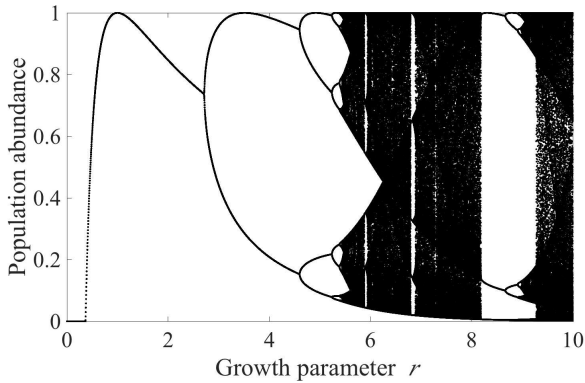}
    \caption{Bifurcation diagram for the normalized Ricker map equation.}
    \label{fig:bifurcation}
\end{figure}

The growth phase is modeled using a {\em normalized Ricker map} $f:[0,\infty) \mapsto [0,1]$ (\cite{Ricker54}), given by 

\begin{equation*}
    f(x) = rxe^{(1-rx)}
\end{equation*}
where $r>0$ is the {\em growth parameter}.  This map exhibits similar behavior to the well-known logistic map, including a period-doubling cascade and chaos for certain values of $r$ (see Figure \ref{fig:bifurcation}).  

The normalized Ricker map is applied patch-wise so that for input population abundance matrix $X_n$, $\bar{X}_n$, where

\begin{equation*}
\bar{X}_n(i,j) = f(X_n(i,j)) 
\end{equation*}

\noindent is the abundance matrix following the growth phase and $n$ is the time index.  

In the dispersal phase, a set fraction of the population in each patch, dictated by the {\em dispersal parameter} $0\leq d \leq 1$, disperses symmetrically into the four nearest neighboring patches on the lattice.  This results in $\bar{\bar{X}}_n$, given input population abundance matrix $X_n$ and $\bar{X}_n$ as defined above, where

\begin{equation*}
    \bar{\bar{X}}_n(i,j)=(1-d)\bar{X}_n(i,j)+\frac{d}{4}\bigg[\bar{X}_n(i-1,j)+\bar{X}_n(i+1,j)+\bar{X}_n(i,j+1)+\bar{X}_n(i,j-1)\bigg].
\end{equation*}
Here, $\bar{X}_n(l, k):=0$ if $l$ and/or $k$ is equal to 1 or $N+1$.

After the dispersal phase, we apply a {\em local extinction threshold} patch-wise, so

\begin{equation*}
{X}_{n+1}(i,j) = \left\{ \begin{array}{cc}
\bar{\bar{X}}_n(i,j) & \mbox{if $\bar{\bar{X}}_n(i,j)>\epsilon$}\\
0 & \mbox{otherwise.}
\end{array}
\right.
\end{equation*}
Therefore, if the population abundance on a patch dips below the local extinction value of $\epsilon$, then that patch experiences a local extinction event resulting in an abundance of zero.  In this way we introduce an allee effect, as patches with low population abundances will be set to zero and will be unsuccessful at reproduction in the following generation.

As we keep the fitness parameter $r$ and dispersal parameter $d$ to be the same across all patches, the spatial domain is assumed homogeneous and there is no preferred directionality to the dispersal.  If part of the population disperses outside of the confines of the $N\times N$ lattice, it is considered lost to the system.  This is an absorbing boundary condition.  Thus, the 2-dimensional domain represents the habitable range of the modeled population, and outside of the domain the species cannot survive due to environmental conditions, resource availability, etc.  

In future work we explore additional mechanisms of dispersal, as well as stochastic growth and dispersal, to make the model more applicable to real-world systems.  However, for this proof-of-concept paper, we choose this simple, yet dynamically rich, model.    

It is natural to visualize a population distribution of this type as a greyscale image where each pixel corresponds to a patch and the greyscale value on that pixel corresponds to the abundance on that patch.  Since we will be focusing on computing topological information on sets, we must first extract a region of interest.  In this work, we focus on the collection of patches that have positive abundance values.  We will call these patches ``occupied'' and color them black.  Since patches with an abundance of ``0'' are already white, this corresponds to converting the greyscale image for the population distribution to a binary image (see Figure~\ref{fig:dispersalexamples} for example greyscale and binary images).  
The black set given by the union of the occupied patches is the set or {\em pattern} we will study.  

\subsection{Computational homology and Betti numbers}
In this work, the key approach we employ for measuring spatial population patterns utilizes cubical homology.  We will focus on computing homological information about the collection of occupied (black) patches.  By construction, these patches live on a cubical lattice and the collection may be specified as a list of closed, two-dimensional squares (cubes) of the form $[i, i+1] \times [j, j+1]$ for some integers $1\leq i, j\leq N$.  That is, for population abundance matrix $X$, the {\em collection of occupied patches}, $\mathcal{X}^+$, is

\begin{equation*}
\mathcal{X}^+:=\{[i, i+1] \times [j, j+1] | \, X(i,j)>0\}.
\end{equation*}

\noindent We note here that a subset of patches may be chosen differently, for example by defining the super level set $\mathcal{X}^\tau:= \{[i, i+1]\times[j, j+1] | X(i,j)>\tau\}$ for some threshold $\tau>0$.  The corresponding {\em population pattern} (also known as the {\em topological realization of $\mathcal{X}^+$}) is 

\begin{equation*}
|\mathcal{X}^+|:=\bigcup_{[i,i+1]\times[j,j+1] \in \mathcal{X}}[i,i+1]\times[j,j+1],
\end{equation*}

\noindent that is, the spatial pattern formed by the collection of occupied patches.

Betti numbers, $\beta_k$, arise in the field of homology as the ranks of the free parts of homology groups. These are computable for cubical sets; see \cite{Kaczynski10} for further explanation.  In what follows, we use the CHomP package (Computational Homology Project software, available at chomp.rutgers.edu) to compute all Betti numbers.  Beyond being computable, what is important for our purposes is the fact that Betti numbers count holes of various dimensions.  For a collection of two-dimensional cubes $\mathcal{X}^+$, $\beta_0 = \beta_0(|\mathcal{X}^+|)$ is the number of connected components ($0$-dimensional holes) in $|\mathcal{X}^+|$, $\beta_1 = \beta_1(|\mathcal{X}|^+)$ is the number of 1-dimensional holes in $|\mathcal{X}^+|$, and all other Betti numbers are $0$.  In our context, $\beta_0$ gives the number of connected regions of populated patches while $\beta_1$ gives the number of enclosed dead (non-populated) regions.  See Figure~\ref{fig:GLS} for example population patterns and their corresponding Betti numbers.



\section{Results}

The following three subsections describe sample results to illustrate this method.  In the first subsection we demonstrate how we use Betti numbers as a low-dimensional metric for interpreting spatiotemporal behavior of the model.  In the second subsection we view an extinction event through the lens of Betti numbers, and in the third subsection we use Betti numbers to characterize the model parameter space.    

\subsection{Examining spatiotemporal dynamics via Betti numbers}
\label{sec:example_ts}
\begin{figure}
    \centering
    \includegraphics[scale =0.7]{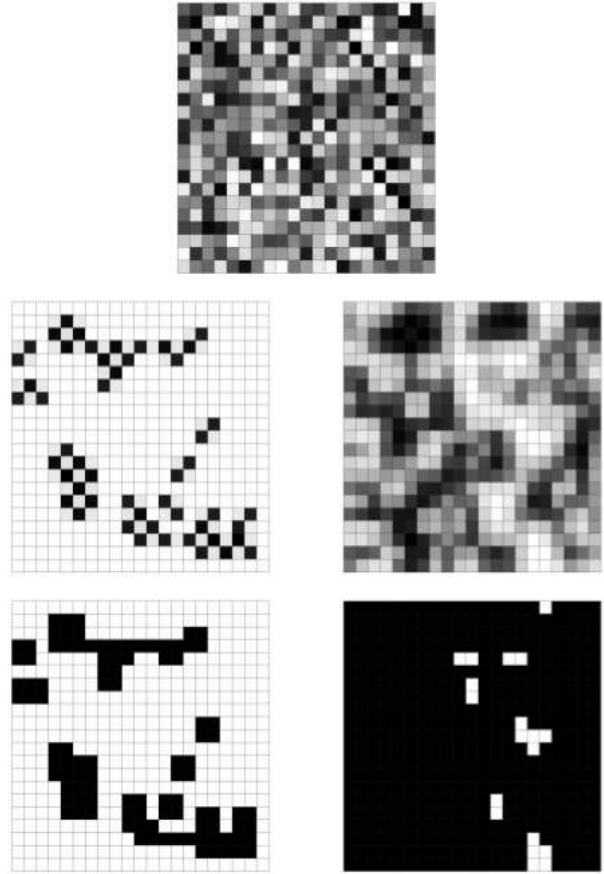}
    \caption{Example population distributions for low dispersal (middle row, left image, $d=0.1$) and high dispersal (middle row, right image, $d=0.5$).  
    Each population distribution is taken from the 100th iterate of a model run with initial condition shown at top.  The bottom row depicts the corresponding (binary) population patterns for the distributions in the middle row.  Betti numbers are computed for the patterns given by the union of the closed, black (occupied) patches.  For the left image, bottom row, $\beta_0 = 7$ and $\beta_1 = 0$.  For the right image, bottom row, $\beta_0 = 1$ and $\beta_1 = 5$.  Patch boundaries are displayed for visualization purposes.}
    \label{fig:dispersalexamples}
\end{figure}

We explore how model dynamics are interpreted through Betti numbers, focusing on two dispersal examples, keeping all other parameters fixed.  The lattice size is set to $N=21$ for easy visualization, the growth parameter ($r$) is set to 8 (within the chaotic range of the normalized Ricker map).  Recalling that the population in a given patch can fluctuate between zero and one, the local extinction threshold ($\epsilon$) is set to 0.06.  The two dispersal parameters used are $d = 0.1$ (low dispersal) and $d = 0.5$ (high dispersal).  The initial condition consists of assigning each patch in the lattice a random value between zero and one, drawn from a uniform distribution.  The initial condition chosen for both trials is depicted in Figure \ref{fig:dispersalexamples} (top row).

The spatial scale of the population distribution is dependent on the dispersal parameter, which we illustrate in Figure \ref{fig:dispersalexamples}.  The greyscale images in the middle row are snapshots of one generation of the population model, out of 100 total saved generations (each population matrix is the 100th iterate).  The spatial structures of the left image are noticeably smaller than the right image, emulating an irregular checkerboard pattern.  The differences in pattern are also reflected in the Betti numbers, where the low dispersal population has more connected components and fewer holes, indicating that the spatial structure is composed of more disconnected regions of populated patches that do not directly communicate via dispersal, at least under one iteration of the map.  In contrast, the high dispersal population has one connected component and many holes, indicating that most of the patches in the lattice are occupied and connected via dispersal.  This is shown in the bottom row of Figure \ref{fig:dispersalexamples}, where the population distributions have been converted to the corresponding (binary) population patterns.

\begin{figure}
    \centering
    \includegraphics[width=\linewidth]{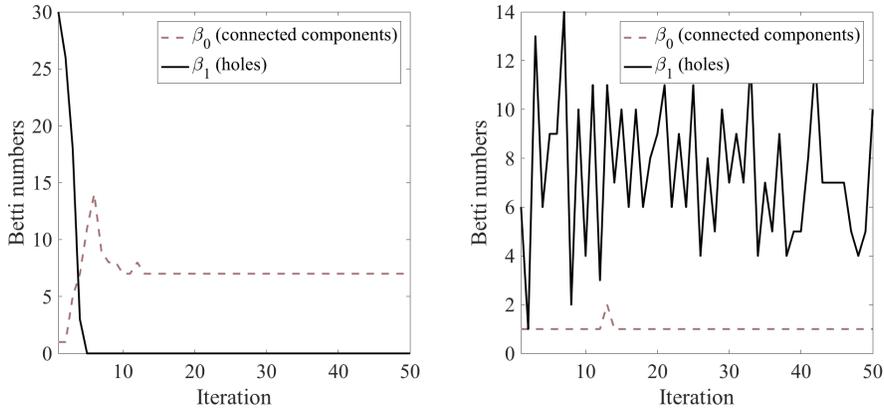}
    \caption{Betti number time series for the parameter combinations illustrated in Figure \ref{fig:dispersalexamples}, low dispersal on the left ($d = 0.1$) and high dispersal on the right ($d = 0.5$).  Via the Betti number time series, we observe that the low dispersal parameter combination appears to achieve steady-state (in terms of Betti numbers), with zero holes and seven connected components.  We refer to this as a {\em topological steady state}.  That is, the topology of the pattern of occupied patches is fixed, although the abundance of individual patches, as well as location and number of occupied patches, may continue to fluctuate.  In contrast, the high dispersal parameter combination does not achieve steady state, but on average maintains a low number of connected components and many holes.  Fluctuating Betti numbers means that the abundance values are necessarily fluctuating as well.}
    \label{fig:timeseriesexample}
\end{figure}

Betti numbers provide a mechanism for quantitatively assessing spatial patterns in a population distribution, and Betti number time series provide a mechanism for assessing the long-term dynamics of a population via its changing spatial patterns.  It must be noted, however, that Betti numbers are a coarse-grain method for observing dynamical changes and information is inevitably lost.  For example, Figure \ref{fig:timeseriesexample} displays the Betti number time series for the parameter combinations illustrated in Figure \ref{fig:dispersalexamples}.  
The low dispersal population appears to achieve steady-state, with three connected components and zero holes.  If, however, we observe the time series of the population distribution itself (non-binary), the individual patches that compose the larger connected components continuously flip between high- and low-density populations (due to the density dependence), thus, the true dynamics of this parameter combination are nearly period-two (there are small fluctuations in the total population, so this parameter combination is not actually period-two and, in fact, does not repeat in 100 iterates).


For the high dispersal parameter combination (Figure \ref{fig:timeseriesexample}, right image), the Betti number time series illustrates the continuously changing spatial structure of the population.  The number of connected components is consistently low, reflecting the fact that most of the patches in the lattice are occupied, while the number of holes remains high, on average, but fluctuates from generation to generation.  Thus, the spatial pattern of this population is continuously changing and does not achieve any type of steady-state.  

\subsection{Critical transitions and Betti numbers}

Observing an extinction event via Betti number time series reveals a characteristic manner in which the population goes extinct.  We use a parameter combination which results in a global extinction event ($d = 0.1$, $r = 8$, $\epsilon= 0.07$, lattice size $N = 21$).  As before, for the initial condition, each patch is assigned a random value between zero and one, drawn from a uniform distribution.  Starting with a random initial condition in which all patches are occupied, there is a rapid increase in the number of connected components and a rapid decrease in the number of holes under iteration of the model.  This signifies the breakup of large connected components into smaller decoupled components as the population undergoes fragmentation.  The number of one-dimensional holes then drops to zero as the number of connected components continues to increase, representing both an increase in the number of connected regions of occupied patches and a topological simplification of these regions.  Thus, regions of occupied patches become increasingly fragmented as they decouple from neighboring components and decrease in size and topological complexity.  Finally, the connected components reach a maximum value before dropping, eventually to zero.  In this final stage, the remaining occupied patches are isolated and receive no additional population from neighboring patches (as the neighboring patches are empty).  The isolated patches cannot sustain their local populations and go extinct.   

\begin{figure}
    \centering
    \includegraphics[width=\linewidth]{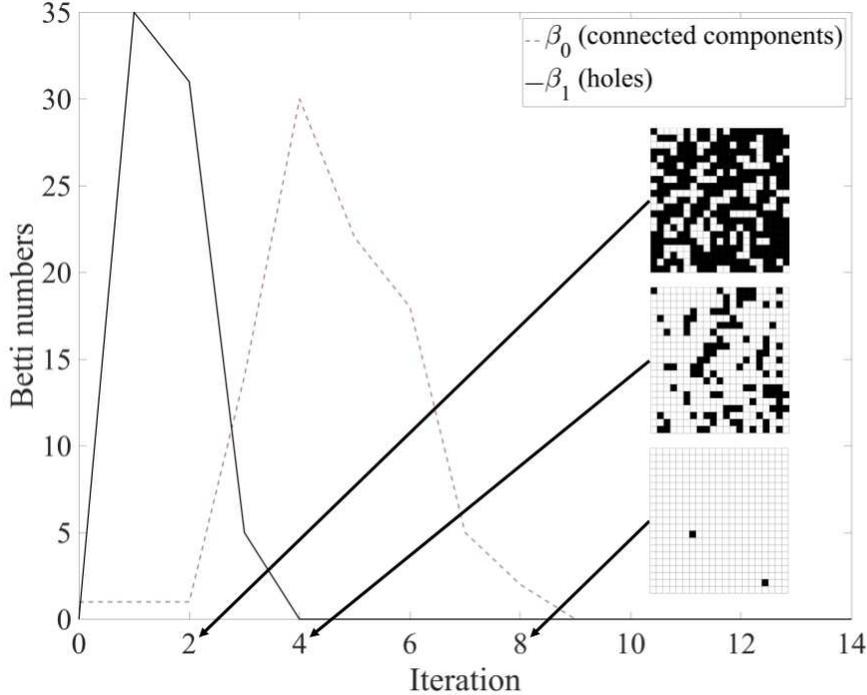}
    \caption{Betti number time series for an extinction event.  Parameters are: $N = 21$, $r = 8$, $d = 0.1$.  Initial condition is random, similar to the initial condition in Figure \ref{fig:dispersalexamples}, local extinction threshold $\epsilon = 0.07$.  En route to a global extinction event, the Betti number time series displays a characteristic shape, where first the number of holes ($\beta_1$) drops to zero while the number of connected components ($\beta_0$) increases (signaling a decoupling of the populated regions), and then connected components drop to zero as the decoupled regions cannot maintain population abundance without immigration from neighboring patches.  The figure contains three snapshots of the population patterns en route to extinction, at iterations (generations) two, four, and eight.}
    \label{fig:extinctionevent}
\end{figure}

We illustrate this general trend in Figure \ref{fig:extinctionevent}, which displays the Betti number time series of an extinction event.  There are three example population patterns taken from three different iterations of the time series. 

\subsection{Characterizing the model parameter space}

We now measure asymptotic dynamics in the model over a large portion of the parameter space, using ``traditional'' methods first and then progressing to Betti numbers.  We first estimate the number of occupied patches typical for a given parameter combination after many iterations of the map.  In order to obtain this information, we run the model using 200 random initial conditions and average the results from the 200 trials.  As before, the initial conditions are constructed by assigning each patch in the lattice a uniformly distributed random number between zero and one.  Information is obtained from the 1000th iterate of each trial.  As illustrated in Figures \ref{fig:timeseriesexample} and \ref{fig:extinctionevent}, model dynamics tend to rapidly settle, so we expect 1000 iterations to be sufficient for estimating long-term dynamical behavior.

\begin{figure}
    \centering
    \includegraphics[scale = 0.2]{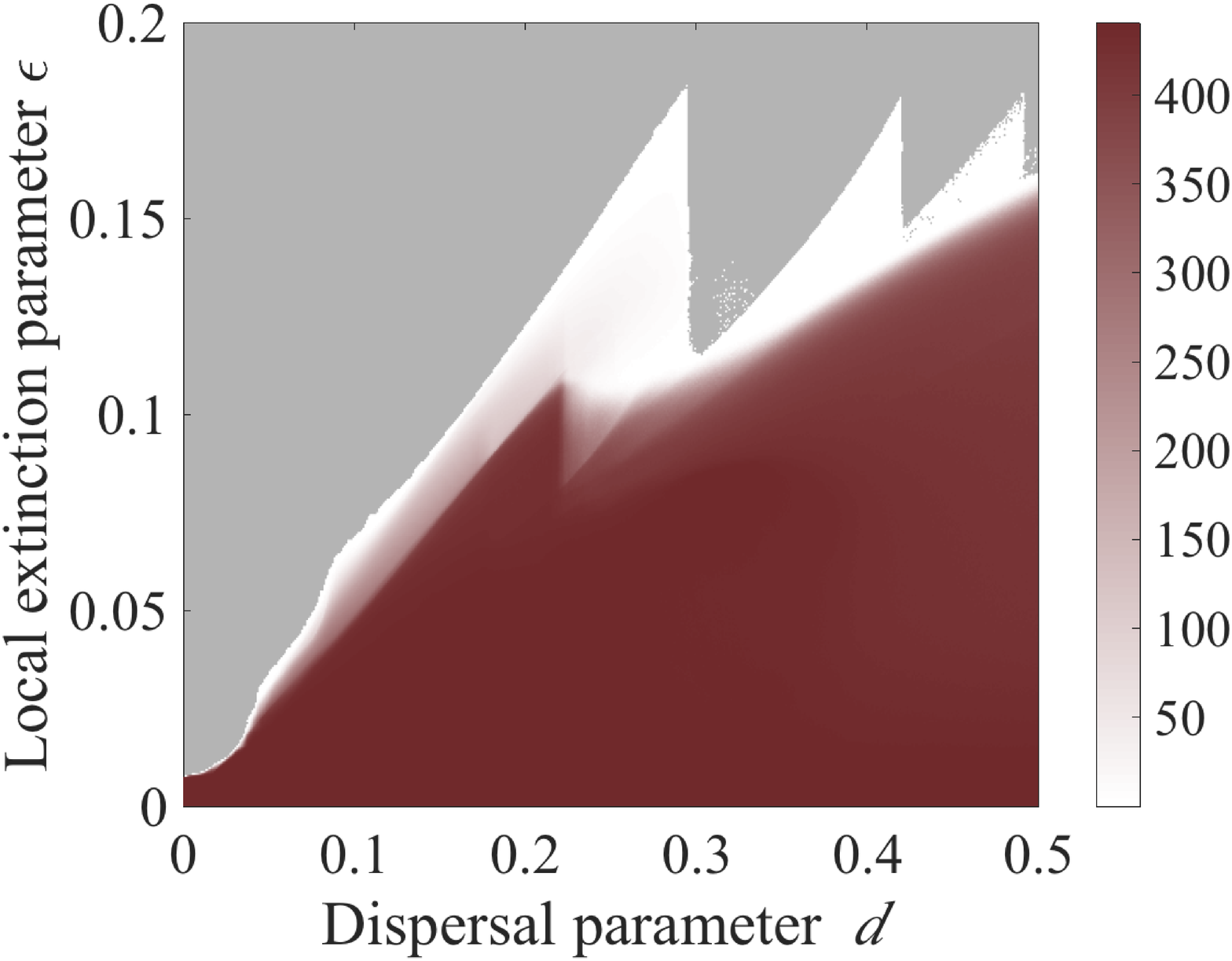}
    \caption{Average number of occupied patches over 200 trials with random initial conditions.  Lattice size $N=21$ (so total number of patches equals 441), growth rate $r=8$.  For lower $\epsilon$ values, the average number of occupied patches is higher, indicating a healthy population.  An average of zero is designated grey (all trials experience global extinction in the grey regions).}
    \label{fig:avgoccupied}
\end{figure}

Figure \ref{fig:avgoccupied} displays the average number of occupied patches over a significant portion of the parameter space in $d$ and $\epsilon$ ($0 \leq d \leq 0.5$ and $0 \leq \epsilon \leq 0.2$).  Growth rate and lattice size are fixed ($r = 8$ and $N=21$).  Lower local extinction ($\epsilon$) values yield a high average number of occupied patches, indicating a healthy population.  For local extinction ($\epsilon$) equal to zero, the average number of occupied patches is equal to 441 (all patches occupied), as the Ricker map does not allow the population in a patch to map to zero.  Large dispersal ($d$) values are able to maintain a high average number of occupied patches over a larger range of $\epsilon$ values.  Regions of grey in the figure indicate an average value of zero, thus, the population experiences global extinction in all trials in these regions. 

\begin{figure}
    \centering
    \includegraphics[scale = 0.2]{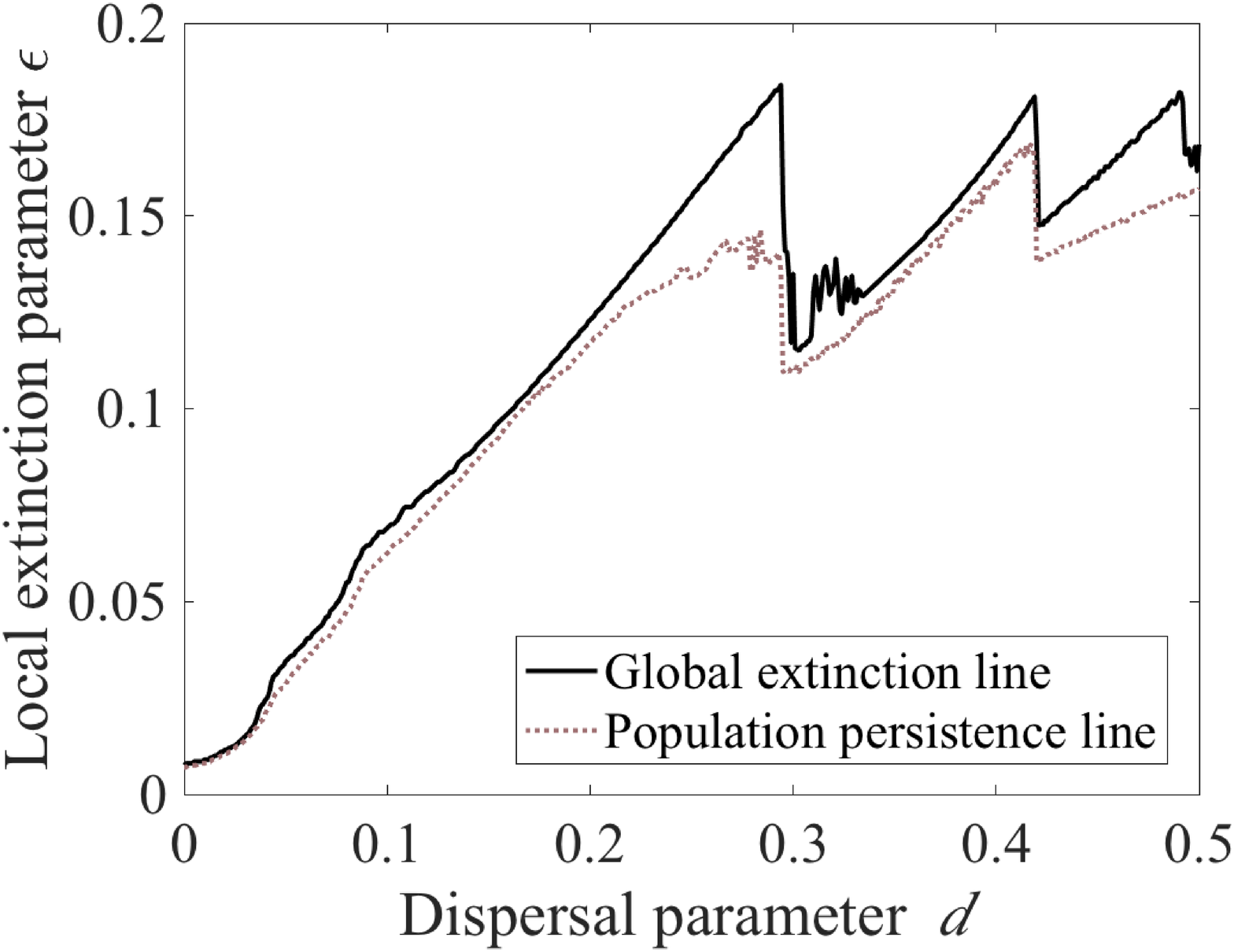}
    \caption{Illustrating regions in the parameter space with persistence versus extinction, based on the local extinction parameter $\epsilon$ versus the dispersal parameter $d$.  Growth rate is $r = 8$, lattice size $N = 21$.  For each parameter combination, 200 trials with random initial conditions were run for 1000 iterations.  Above the solid black line, all 200 trials experience global extinction within 1000 iterations.  Below the dotted red line, all 200 trials persist over 1000 iterations.}
    \label{fig:aliveordead}
\end{figure}

To better define the ``sawtooth'' boundary between the global extinction (grey) region and the rest of the parameter space in Figure \ref{fig:avgoccupied}, we determine the {\em global extinction line}, i.e., the smallest local extinction parameter $\epsilon$ for a given dispersal $d$ which leads to global extinction for all 200 trials.  As before, each trial is iterated 1000 times.  We additionally determine the largest possible $\epsilon$, for a given $d$, in which all 200 trials are still alive after 1000 iterations (i.e., the {\em population persistence line}).  Figure \ref{fig:aliveordead} displays the global extinction and population persistence lines for the tested parameter space.

\begin{figure}
    \centering
    \includegraphics[width=\linewidth]{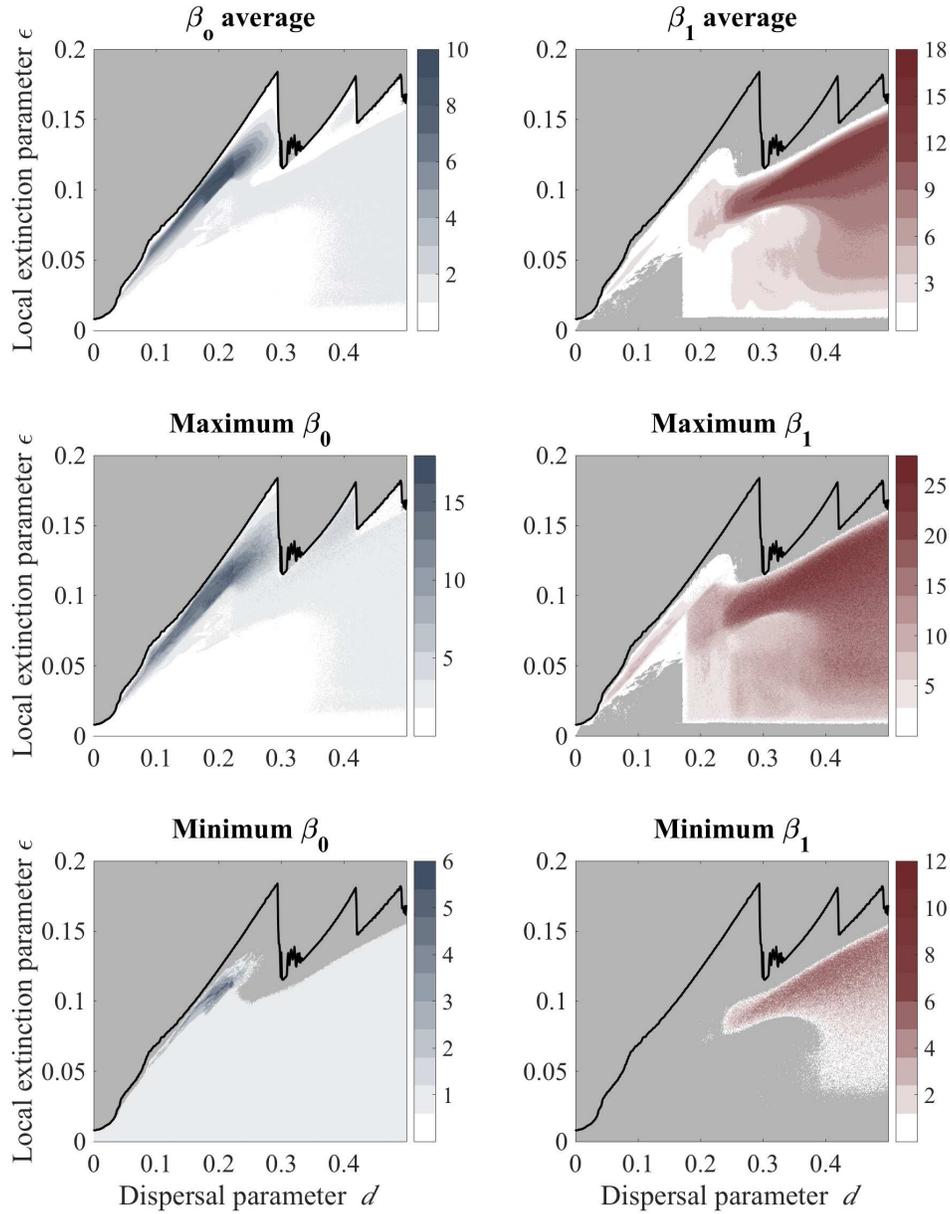}
    \caption{Averages, maxima, and minima for $\beta_0$ (left) and $\beta_1$ (right).  For each parameter combination, 200 trials with random initial conditions were run for 1000 iterations each, information obtained from the last iteration.  Top row: Betti number averages over the 200 trials.  Middle row: maximum $\beta_0$ and $\beta_1$ values observed over 200 trials.  Bottom row: minimum $\beta_0$ and $\beta_1$ values observed over 200 trials.  The growth rate is $r = 8$, lattice size $N = 21$.  Grey indicates a value of zero.  The black solid line indicates, for a given dispersal $d$, the smallest $\epsilon$ for which all trials go extinct.  Please note the different scales of each colorbar.}
    \label{fig:avgmaxmin}
\end{figure}

To obtain spatial information about the population dynamics below the global extinction line, we calculate average $\beta_0$ and $\beta_1$ values over the parameter space used in Figures \ref{fig:avgoccupied} and \ref{fig:aliveordead}.  As before, $\beta_0$ and $\beta_1$ values are obtained from the 1000th iterates of each trial, with 200 total trials.  Figure \ref{fig:avgmaxmin} displays the $\beta_0$ and $\beta_1$ averages (top row), maxima (middle row), and minima (bottom row).  The maxima display the highest $\beta_0$ and $\beta_1$ values observed for each parameter combination over 200 trials, and the minima display the smallest $\beta_0$ and $\beta_1$ values observed for each parameter combination over 200 trials.  As in Figure \ref{fig:aliveordead}, the solid black line indicates, for a given $d$, the smallest $\epsilon$ for which all 200 trials go extinct, and is included as a reference to easily compare across figures.  

We observe areas of higher $\beta_0$ averages near the global extinction line, which rapidly drop to zero as they cross the line.  For example, using $d = 0.2$ and moving vertically upward in the $\beta_0$ averages plot, when $\epsilon$ is low, most patches in the lattice are occupied, and $\beta_0$ is small.  As $\epsilon$ increases, $\beta_0$ also increases as the large regions of occupied patches are broken up into smaller disconnected components.  As the limit of global extinction approaches, the number of disconnected components rapidly drops to zero.  We observe that the $\beta_1$ average is also lower for small $\epsilon$, as most patches are occupied and thus there are few holes.  As $\epsilon$ increases and more patches experience local extinction, more holes are created ($\beta_1$ increases).  Finally, as the global extinction limit is reached, the population consists exclusively of disconnected small components, and the number of holes drops to zero.  As in the global extinction example above, this is a topological simplification of connected regions of occupied patches.   

\begin{figure}
    \centering
    \includegraphics[width=\linewidth]{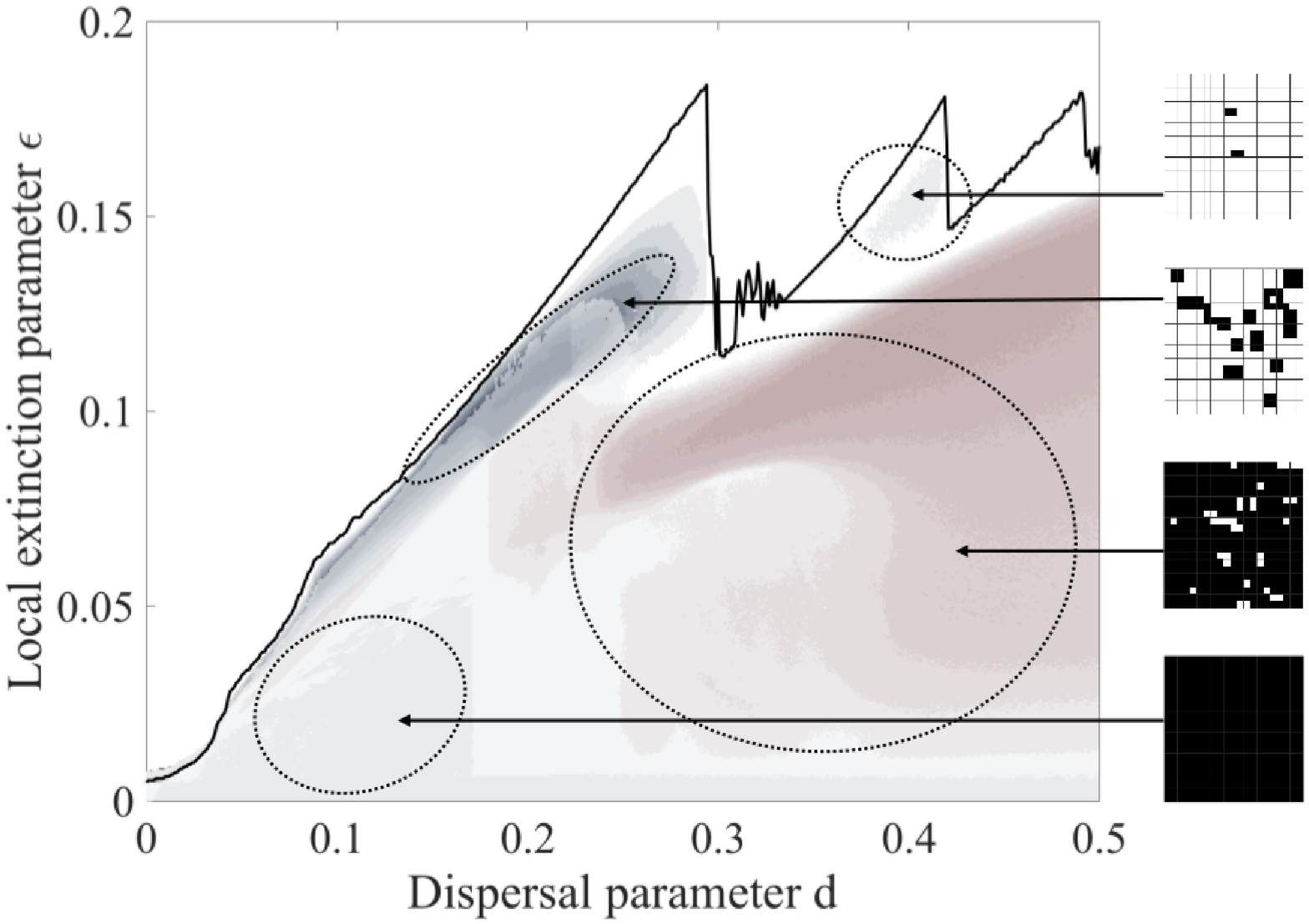}
    \caption{Four example spatial patterns from different regions of the parameter space.  $\beta_0$ (blue) and $\beta_1$ (red) values are indicated by the transparent colored regions, overlapping the two plots from the top row of Figure \ref{fig:avgmaxmin}.  As before, the black solid line indicates the parameter combination for which all 200 trials go extinct.  From bottom to top, (1) regions of $\beta_0 =1$ and $\beta_1 =0$ imply a continuous population (one large connected component), (2) regions of $\beta_0$ low and $\beta_1$ high imply a ``swiss cheese'' population, (3) regions of $\beta_0$ and $\beta_1$ high/nonzero imply multiple disconnected subpopulations, (4) regions of $\beta_0 > 1$ and $\beta_1 = 0$ imply isolated islands of populations containing no holes.}
    \label{fig:betti1betti0combo}
\end{figure}

We can characterize patterns by type based on $\beta_0$ and $\beta_1$ values in different regions of the parameter space.  Figure \ref{fig:betti1betti0combo} illustrates several example patterns.  The average $\beta_0$ and $\beta_1$ plots from the top row of Figure \ref{fig:avgmaxmin} have been overlapped to provide a guide for regions of high $\beta_0$ versus high $\beta_1$.  The four example regions are not meant to provide a full picture of the parameter space or provide the exact boundary of each region.  The four spatial patterns illustrate, from bottom to top, a continuous population ($\beta_0=1$, $\beta_1=0$), a ``swiss cheese'' population ($\beta_0$ low, $\beta_1$ high), a population comprised of multiple disconnected subpopulations with some holes ($\beta_0$ and $\beta_1$ high/nonzero), and a population consisting solely of isolated islands ($\beta_0$ high, $\beta_1=0$).  Figure \ref{fig:betti1betti0combo} additionally illustrates the shortcomings of relying on $\beta_0$ and $\beta_1$ alone.  A continuous population (bottom spatial distribution) and a population with a single connected component both yield $\beta_0=1$ and $\beta_1=0$.

\section{Discussion}
Using a simple deterministic population model, we illustrate how Betti numbers are a useful metric for characterizing spatial patterns in a high-dimensional system.  Calculation of $\beta_0$ and $\beta_1$ allows us to reduce the dimensionality of the system, while still retaining important information about the spatial patterns of the population.  The characteristic changes in $\beta_0$ and $\beta_1$ over time en route to global extinction suggest promising applications of Betti numbers as a tool in the prediction of critical transitions.  Betti number time series can also be utilized to assess the coarse-grain dynamics and stability of a population over time, which may be more relevant to ecologists and managers than assessing the population dynamics in a mathematically rigorous/higher dimensional way, e.g., in Figure \ref{fig:timeseriesexample} the left Betti number time series appears stable but the population is quasi-periodic (nearly period-2).  

This work also illustrates the shortcomings of relying solely on information about $\beta_0$ and $\beta_1$ to assess the health or predict the dynamics of a population, particularly when attempting to ascertain such information from a single point in time.  For example, a population distribution with $\beta_0 =1$ and $\beta_1 =0$ can describe either a single connected component in an otherwise empty lattice, or a contiguous population in which each patch in the lattice is occupied.  Either of these population distributions may appear stable over time with respect to their Betti number time series, but one may be more susceptible to extinction than the other.  Additional information about the system is required in order to assess the population.  We ideally envision Betti number time series being used as an additional tool in the critical transition prediction toolbox, in addition to the techniques mentioned previously (e.g., \cite{Scheffer09}).       

While the deterministic growth and symmetric dispersal model utilized here is a convenient model for illustrating the dynamics of a spatially distributed population and the interpretation of said dynamics via Betti numbers, the model does not provide sufficient ecological realism for direct application to many real-world populations.  
However, the coupled patch lattice model offers a high level of flexibility and is adaptable to a variety of local growth and population dispersal scenarios.  The tools presented here for measuring resulting population patterns remain applicable, and it would be interesting to use them to study more ecologically realistic models, as well as GIS and other digital images of population patterns.

For each of the figures above, we chose a random initial condition in which each patch in the lattice is assigned a random value between 0 and 1 (recalling that the normalized Ricker model maps the population exclusively to [0,1]).  Using this random initial condition, we constructed figures of the parameter space which exhibit a characteristic ``sawtooth'' shape with complicated structure (Figures \ref{fig:avgoccupied}, \ref{fig:aliveordead}, \ref{fig:avgmaxmin}, and \ref{fig:betti1betti0combo}).  We find that the structure of these figures is echoed in alternate initial conditions.  Using the same parameter space as Figures \ref{fig:avgoccupied} through \ref{fig:betti1betti0combo}, we ran the model using a bivariate Gaussian distribution as the initial condition, where the Gaussian is centered in the center of the lattice, i.e., the population is more concentrated in the center.  The Gaussian initial condition also displays the sawtooth pattern with similar locations of high $\beta_0$ and high $\beta_1$ regions, as in Figure \ref{fig:avgmaxmin}.  Dramatically changing the initial condition, however, dramatically alters the figures.  Using, e.g., an initial condition that consists of an empty lattice except for one occupied patch in the corner (invasion scenario), the complicated structures are absent.  The invading population either successfully spreads across the lattice or dies, and so the structures in Figures \ref{fig:avgoccupied} through \ref{fig:betti1betti0combo} that arise from using an initial condition with an established population (with spatial variability) are absent.  

\begin{figure}
    \centering
    \includegraphics[width=\linewidth]{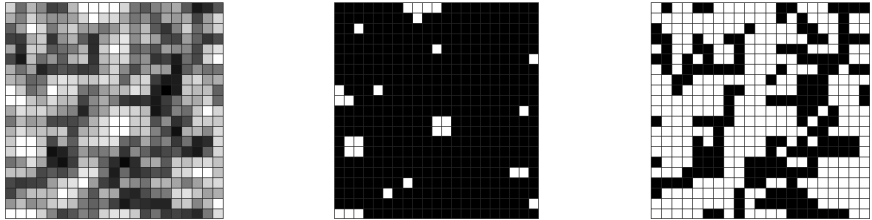}
    \caption{The choice of global thresholding value affects the topology of the spatial pattern.  Moving from left to right: (1) original greyscale image of the population distribution, (2) the population pattern processed with a global thresholding value of 0.05, (3) the population pattern processed with a global thresholding value of 0.5.  For image (2), $\beta_0 =1$ and $\beta_1 =9$, for image (3), $\beta_0 =10$ and $\beta_1 =7$.  Parameters are $d = 0.45$, $\epsilon = 0.1$, $r = 8$, lattice size $N=21$.}
    \label{fig:thresholding}
\end{figure}

A global thresholding value of zero was used for all of the work presented here.  In other words, for binary processing of population distribution images, a patch in the lattice was classified as either unoccupied (value of zero) or occupied (value greater than zero).  For tracking a global extinction event this choice is natural, as we are concerned with the transition from populated to globally unpopulated.  However, for different types of critical transitions or for data analysis, a global thresholding value of zero may not be the most natural choice.  Any choice of global thresholding value can seem somewhat arbitrary, and the value chosen may dramatically impact the topological features of the resulting population pattern.  For example, Figure \ref{fig:thresholding} displays the same population distribution processed with two different global thresholding values.  The leftmost image shows the population distribution, $X$, in greyscale, the middle image shows, $|\mathcal{X}^{0.05}|$, the population processed with a global thresholding value of 0.05, and the right image shows, $|\mathcal{X}^{0.5}|$, the same population processed with a global thresholding value of 0.5.  The topological features of the two patterns differ dramatically.  To avoid this complication, in future work we will employ cubical persistent homology (see e.g., \cite{Mischaikow13} and references therein), which measures topological features across different thresholding values for a given population distribution, tracking the appearance and disappearance of features as the threshold is varied.  

The potential uses of topological analysis in spatial ecology are numerous and diverse.  We have shown how Betti numbers can be used to exhibit characteristic changes occurring in a population during a critical transition, characterize spatial patterns in a model, and reduce the dimensionality of a high-dimensional system.  Betti numbers provide important spatial information that is usually absent in other low dimensional system measurements, such as total population abundance.  Thus, we see it as a powerful tool for understanding and predicting high dimensional spatially explicit systems.  As the dynamics of ecological systems become more volatile due to global climate change, and as we gain the ability to analyze larger and larger spatial data sets with increased computing power, we will require additional methods of analysis.  We believe Betti numbers can serve as an important tool to aid in the understanding of dynamically complex, high dimensional spatial systems.  

\section{Acknowledgements}
S. Day and L. Storch would like to acknowledge the work of B. Holman (\cite{Holman11}), whose undergraduate thesis provided preliminary results in this direction.  S. Day's research was partially sponsored by the Army Research Office and was accomplished under Grant Number W911NF-18-1-0306. The views and conclusions contained in this document are those of the authors and should not be interpreted as representing the official policies, either expressed or implied, of the Army Research Office or the U.S. Government. The U.S. Government is authorized to reproduce and distribute reprints for Government purposes notwithstanding any copyright notation herein.

\bibliographystyle{amsplain}
\bibliography{extinctioneventsmanuscript}

\end{document}